\documentclass[conference]{IEEEtran}
\IEEEoverridecommandlockouts

\usepackage[T1]{fontenc}
\usepackage{newtxtext}
\usepackage{amsmath,amssymb,amsfonts}
\usepackage{graphicx}
\usepackage{booktabs}
\usepackage{array}
\usepackage{makecell}
\usepackage{xcolor}
\usepackage{url}
\usepackage{cite}

\begin{document}

\title{Residual Community Prototypes Under-Reject Held-Out Malware Families in FCG-MFD}

\author{
\IEEEauthorblockN{1\textsuperscript{st} Junru Zhu}
\IEEEauthorblockA{\textit{Independent Researcher}\\
Seattle, WA, USA\\
junru.zhuu@gmail.com}
\and
\IEEEauthorblockN{2\textsuperscript{nd} Yixin Yang}
\IEEEauthorblockA{\textit{Independent Researcher}\\
New York, NY, USA\\
vicki\_yang@chiefgroup.com.hk}
\and
\IEEEauthorblockN{3\textsuperscript{rd} Xiaoqing Ding}
\IEEEauthorblockA{\textit{University of Chicago}\\
Chicago, IL, USA\\
alexading@uchicago.edu}
\and
\IEEEauthorblockN{4\textsuperscript{th} Ruoyu Qi}
\IEEEauthorblockA{\textit{Independent Researcher}\\
Charlotte, NC, USA\\
dxfreedom94@gmail.com}
}

\maketitle

\begin{abstract}
Open-set malware-family recognition must classify known families while
rejecting families absent from training. We test whether Louvain-community
summaries add rejection information beyond a graph neural network embedding
and dimension-matched generic topology. The study uses a deduplicated,
conflict-audited FCG-MFD corpus, five held-out families, and three
optimization seeds. Community features are residualized against generic
topology using known-family training data before nearest-prototype scoring.
Residual community does not produce stable held-out-family rejection.
Ranking effects reverse across families, the false-positive rate at 95
percent unknown recall worsens for every held-out family, and a
validation-fitted threshold rejects only 4.48 percent of unknown samples.
Accepted-known macro F1 improves in every family, but with five independent
family units the exact two-sided sign-flip p-value is 0.0625, the smallest
attainable value. The score remains associated with graph scale, while simple
classifier uncertainty performs better on ranking, high-recall rejection,
and OSCR. In this GIN/FCG-MFD setting, community-enriched prototypes change
known-class geometry without creating a stable unknown margin. Graph
open-set evaluations should pair structural features with matched topology
controls, operational thresholds, and held-out-family analysis.
\end{abstract}


\section{Introduction}

Malware-family classification supports the reuse of analysis across related
samples, but deployed systems also encounter families absent from training.
Function-call graphs (FCGs) represent program structure, and graph neural
networks (GNNs) can learn family representations from these
graphs~\cite{bilot2024survey,ling2022malgraph}. Closed-set accuracy alone
does not measure whether such a representation can recognize when a sample
falls outside its known-family support.

Open-set recognition (OSR) adds this rejection requirement. Activation-space
models and metric-learning methods assign unknown scores from the relation
between a test representation and known-class
support~\cite{bendale2016openmax,hassen2020open}. Malware studies have
applied FCG transformations, learned distances, and contrastive
representations to unknown or shifted
samples~\cite{jia2022fcgosr,yang2021cade,wu2024contrastive}. These methods
make the representation central to rejection: an added graph descriptor is
useful only when it separates unseen families beyond information already
present in the embedding and generic graph topology.

Community organization is a candidate descriptor because modularity,
community-size distributions, and cross-community connectivity summarize
mesoscopic structure. The same summaries, however, vary with graph size,
density, and degree. A prototype can therefore respond to structural scale
rather than family novelty. This confounding is especially important under
held-out-family evaluation, where the size distribution itself may shift.
A controlled test must compare community information with generic topology
and examine operating thresholds, scale dependence, and family-specific
effects rather than relying on a pooled ranking metric.

We conduct this test on a cleaned FCG-MFD corpus~\cite{hadi2025fcgmfd}.
Exact graph conflicts are removed as groups, cross-label
Weisfeiler--Lehman signatures are excluded, and same-label signatures are
collapsed before splitting. Five malware families are held out in turn, with
three optimization seeds per family. A common multiclass GIN provides the
graph embedding. The primary comparison augments this embedding with either
an eight-dimensional generic-topology block or an eight-dimensional
community block residualized against generic topology on known-family
training data.

In this GIN/FCG-MFD setting, the embedding plus residual-community prototype
does not establish reliable unknown-family rejection. Its mean AUROC and
AUPRC changes are small relative
to the dimension-matched generic-topology control. FPR@95TPR is worse across
the held-out-family means, mean OSCR does not improve, and the
validation-fitted threshold accepts 95.52\% of held-out-family samples. The
mean known-family macro F1 is 0.0082 higher, but family-clustered bootstrap
uncertainty includes zero. Scores remain associated with graph scale, and
effects reverse between winwebsec and WannaCry\_Worm. Together, these
observations locate the failure in the tested representation and operating
regime rather than supporting a general benefit from community features.

Our contributions are:
\begin{itemize}
    \item a leakage-audited FCG-MFD protocol with deterministic graph-group
    splitting, five held-out families, three optimization seeds, and retained
    sample-level scores;
    \item a dimension-matched comparison that tests whether residual
    community structure adds unknown-family information beyond a learned
    embedding and generic topology;
    \item a threshold-, scale-, and family-aware negative result showing that
    the tested structural prototype under-rejects held-out families and
    trails classifier-uncertainty baselines in this GIN/FCG-MFD setting.
\end{itemize}

\section{Related Work}

\subsection{Graph Representations for Malware Families}

Graph-based malware learning represents programs through call and control
flow rather than only flat feature vectors~\cite{bilot2024survey}. Function
call graph embeddings have been used for Android malware detection and family
categorization~\cite{xu2021detecting}, while hierarchical GNNs combine
inter-function call graphs with intra-function control-flow
graphs~\cite{ling2022malgraph}. Contrastive objectives have also been applied
to malware familial classification~\cite{wu2024contrastive}. FCG-MFD extends
this line with a function-call-graph benchmark spanning malware
families~\cite{hadi2025fcgmfd}. These studies establish graph encoders for
closed-set prediction. Our study instead holds a common encoder fixed and
tests whether an added community block supports rejection of a family absent
from training.

\subsection{Open-Set and Shift-Aware Malware Recognition}

General OSR methods derive unknown scores from class activation tails or
distance in a learned representation
space~\cite{bendale2016openmax,hassen2020open}. The closest graph-specific
malware study learns representations from transformed FCGs and applies
distance-based rejection to unseen families~\cite{jia2022fcgosr}. CADE
similarly learns a contrastive distance for detecting and explaining
individual drift samples~\cite{yang2021cade}. CNS-Net synthesizes
conservative novelty examples for broader open-set malware recognition
~\cite{guo2023cnsnet}. These methods establish rejection strategies; our
novelty claim is limited to the matched structural control and diagnostic
failure analysis, not a new rejection rule.

Malware reliability research has also examined predictive uncertainty and
selective classification under dataset, adversarial, and temporal
shift~\cite{li2021uncertainty,li2024malcertain,herzog2025selective}.
TESSERACT further shows that spatial and temporal split choices can bias
malware-classification evaluation~\cite{pendlebury2019tesseract}. These lines
motivate classifier-uncertainty baselines, held-out-family reporting, and
operating-point metrics. Our comparison asks a narrower structural question:
after an embedding and generic topology are available, does residual
community information improve family rejection?

\subsection{Community Structure and Scale Controls}

Louvain detects communities by optimizing modularity through local moves and
graph coarsening~\cite{blondel2008louvain}. Modularity optimization has a
resolution limit: the smallest recoverable modules depend on total network
size and their interconnection pattern~\cite{fortunato2007resolution}.
Community count, modularity, and size-distribution summaries are therefore
not independent of graph scale. Under held-out-family shift, these quantities
can change prototype distance without encoding family-specific novelty.

We isolate this ambiguity by residualizing an eight-dimensional community
block against generic topology on known-family training data and comparing it
with an eight-dimensional generic-topology control. Score--size correlations
and size-stratified metrics then test whether rejection behavior remains
associated with structural scale.

\section{Method}

\subsection{Study Design and Leakage Controls}

We use the FCG-MFD corpus~\cite{hadi2025fcgmfd} to evaluate whether community
structure adds open-set information beyond an FCG embedding and generic
topology. The source archive contains 46,714 graph rows. Cleaning precedes
all splitting: empty and edgeless graphs are removed, exact canonical graphs
with conflicting family labels are dropped as groups, cross-label
Weisfeiler--Lehman signatures are excluded, and same-label signatures are
collapsed to one representative. Benign graphs are excluded because the task
is malware-family recognition.

The resulting model pool contains 20,870 structural signatures across 32
labels. Twenty-three malware families meet the 100-signature modeling
threshold. Five families selected before model outcomes are held out in turn:
BazaLoader (859 graphs), Mirai Botnet (1,212), Zeus Trojan (1,045),
winwebsec (2,996), and WannaCry\_Worm (2,714). Each fold assigns the entire
held-out family to unknown testing. Within every remaining family, a
deterministic graph-group split assigns 70\% to training, 15\% to validation,
and 15\% to known testing. The partitions are fixed across seeds.

The unknown family is absent from encoder training, checkpoint selection,
feature scaling, residualization, prototype fitting, and threshold selection.
The encoder and structural score models use known-family training data;
checkpoint selection and rejection thresholds use known-family validation
data. The unknown-family split is accessed only for final evaluation.

\subsection{Graph and Structural Representations}

Directed call edges define five node features: log in-degree, log out-degree,
log total degree, and source and sink indicators. Message passing uses the
symmetrized edge set. A three-layer GIN~\cite{xu2019gin} with 64 hidden units
and a 64-dimensional graph embedding applies global mean and max pooling.
Training uses class-weighted cross entropy, dropout 0.2, Adam with learning
rate $10^{-3}$ and weight decay $10^{-4}$, batches of 64 graphs, and early
stopping with patience 12 within an 80-epoch cap. The selected checkpoint
maximizes known-validation macro F1, with negative log-likelihood as the
tie-breaker. Optimization seeds are 7, 23, and 47.

Structural covariates are computed on the undirected simple graph induced by
the FCG. The eight-dimensional generic-topology block contains log node and
directed-edge counts, undirected density, mean and standard deviation of
degree, maximum-degree ratio, component ratio, and transitivity. Louvain
community detection uses resolution 1 and seed 0
~\cite{blondel2008louvain}. Its eight-dimensional block contains modularity,
log community count, community-count ratio, normalized community-size
entropy, maximum-community ratio, intra-community edge ratio, bridge-node
ratio, and size-weighted within-community density.

\subsection{Dimension-Matched Prototype Scores}

Let $z(x)\in\mathbb{R}^{64}$ denote the GIN embedding, $u(x)\in\mathbb{R}^8$
the generic block, and $q(x)\in\mathbb{R}^8$ the community block. Training-set
standardizers $S_u$ and $S_q$ put the structural coordinates on comparable
scales. A multi-output ridge model $g$ with penalty 1 predicts standardized
community structure from standardized generic topology. The residual is
\begin{equation}
r(x)=S_q(q(x))-g\!\left(S_u(u(x))\right).
\end{equation}
Thus, residualization removes the linear component predictable from generic
topology in known-family training data; it does not impose independence under
held-out-family shift.

For a feature space with blocks $h_b(x)$ of dimensions $d_b$, each block is
standardized on training data and scaled by $1/\sqrt{d_b}$ before
concatenation. The prototype and unknown score are
\begin{equation}
\mu_c=\frac{1}{|T_c|}\sum_{x_i\in T_c}\bar h(x_i),\qquad
s(x)=\min_c\left\|\bar h(x)-\mu_c\right\|_2 ,
\end{equation}
where $T_c$ is the training set for known family $c$. The scaling prevents
the 64-dimensional embedding from dominating an eight-dimensional
structural block solely through dimension. The primary comparison uses
$\bar h=[z,u]$ for the generic-topology control and $\bar h=[z,r]$ for the
residual-community score.

We additionally evaluate MSP using $1-\max_c p(c\mid x)$, normalized
predictive entropy, energy~\cite{liu2020energy}, and prototypes built from
the embedding, generic block, or raw community block alone. Larger scores
always indicate stronger evidence for rejection. For each method, the
rejection threshold is the empirical 95th percentile of its known-validation
scores, targeting 95\% known-family acceptance without using unknown samples.

\subsection{Metrics and Family-Clustered Analysis}

Unknown AUROC and AUPRC measure ranking. FPR@95TPR reports the known-sample
false-positive rate when unknown recall reaches 95\%; lower is better. OSCR
combines correct known classification with unknown rejection
~\cite{dhamija2018reducing}. We also report known-family macro F1, known
acceptance, and unknown rejection at the validation-fitted threshold.

Each metric is computed within a held-out-family and optimization-seed cell.
Seeds are repeated optimization runs, whereas the five held-out families are
the units of generalization. For the primary comparison, we compute paired
method differences within each cell, average the three seeds within each
family, and obtain 95\% intervals by resampling the five family means 10,000
times with replacement~\cite{efron1979bootstrap}. Oriented effects are
positive when residual community is better; FPR@95TPR differences are
sign-reversed. Absolute means over the 15 cells give each family equal weight
because every family has three seeds. Because only five family units are
available, percentile intervals are descriptive. We also enumerate all 32
sign assignments to the five family means for an exact two-sided sign-flip
test; the smallest attainable p-value is 0.0625.

\section{Experiments and Analysis}

\subsection{Classifier Uncertainty Is the Stronger Baseline}

Table~\ref{tab:absolute} reports means over the 15 family--seed cells. The
three classifier-uncertainty scores outperform both concatenated prototype
scores on unknown AUROC and OSCR. Entropy attains the highest AUROC (0.6149)
and lowest FPR@95TPR (0.7101), while MSP attains the highest OSCR (0.2839).
The residual-community prototype reaches 0.5319 AUROC, 0.9160 FPR@95TPR,
and 0.1725 OSCR. Its ranking is therefore weaker than simple uncertainty,
and its high FPR shows that reaching 95\% unknown recall requires rejecting
most known samples.

\begin{table}[t]
\caption{Mean performance over five held-out families and three seeds.
Classifier uncertainty is stronger than the concatenated prototypes on
unknown ranking and OSCR. Higher is better except FPR@95TPR.}
\label{tab:absolute}
\centering
\footnotesize
\setlength{\tabcolsep}{1.2pt}
\begin{tabular}{lrrrrr}
\toprule
Method & AUROC & AUPRC & FPR95 & OSCR & F1\\
\midrule
MSP & 0.6013 & 0.4191 & 0.7152 & \textbf{0.2839} & \textbf{0.2479}\\
Entropy & \textbf{0.6149} & 0.4262 & \textbf{0.7101} & 0.2822 & \textbf{0.2479}\\
Energy & 0.6072 & \textbf{0.4322} & 0.7742 & 0.2558 & \textbf{0.2479}\\
Embed.+generic & 0.5196 & 0.3719 & 0.8941 & 0.1814 & 0.2061\\
Embed.+resid. & 0.5319 & 0.3820 & 0.9160 & 0.1725 & 0.2143\\
\bottomrule
\end{tabular}
\end{table}

\subsection{Residual Community Does Not Improve Family-Level Rejection}

Table~\ref{tab:paired} gives the primary paired comparison at the
held-out-family level. Residual community changes unknown AUROC by +0.0123
and AUPRC by +0.0101, but both family-clustered intervals include zero.
OSCR changes by $-0.0090$ and improves for only one of five family means.
The clearest adverse result is FPR@95TPR: its oriented effect is $-0.0219$,
the interval remains below zero, and all five held-out families favor the
generic-topology control. Its exact family-level sign-flip p-value is 0.0625,
so the unanimous direction is suggestive rather than conventionally
significant.

\begin{table}[t]
\caption{Residual-community versus generic-topology paired effects.
Descriptive intervals resample the five seed-averaged held-out-family effects
10,000 times. Positive values favor residual community; FPR@95TPR is
sign-reversed.}
\label{tab:paired}
\centering
\footnotesize
\setlength{\tabcolsep}{1.0pt}
\begin{tabular}{lrrr}
\toprule
Metric & Mean & 95\% interval & Fam. +\\
\midrule
Unknown AUROC & +0.0123 & [$-$0.0408, +0.0686] & 3/5\\
Unknown AUPRC & +0.0101 & [$-$0.0241, +0.0448] & 4/5\\
FPR@95TPR & $-$0.0219 & [$-$0.0371, $-$0.0069] & 0/5\\
OSCR & $-$0.0090 & [$-$0.0302, +0.0162] & 1/5\\
Known F1 & +0.0082 & [$-$0.0025, +0.0177] & 4/5\\
Accepted F1 & +0.0138 & [+0.0088, +0.0196] & 5/5\\
\bottomrule
\end{tabular}
\end{table}

\begin{figure*}[t]
\centering
\includegraphics[width=\textwidth]{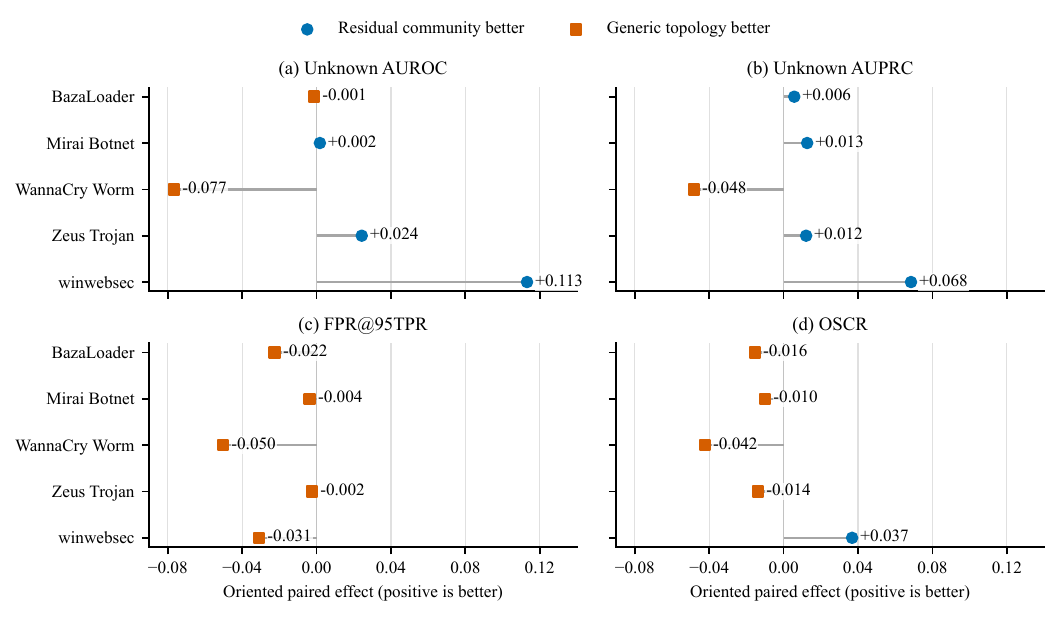}
\caption{Seed-averaged paired effects by held-out family for residual
community versus generic topology. FPR@95TPR is sign-reversed, so rightward
values always favor residual community. Ranking effects reverse across
families, while FPR@95TPR is worse for every held-out family.}
\label{fig:primary}
\end{figure*}

Figure~\ref{fig:primary} shows why the mean ranking changes are inconclusive.
The AUROC effect ranges from $-0.0767$ for WannaCry\_Worm to +0.1132 for
winwebsec. AUPRC exhibits the same reversal, and winwebsec is the only family
with a positive OSCR effect. Averaging these opposing cases produces a small
positive AUROC mean without a family-general improvement.

Classification and rejection also separate. Known-family macro F1 changes
from 0.2061 to 0.2143, but its family-clustered interval includes zero. Among
accepted known samples, macro F1 increases by 0.0138 with a positive
descriptive interval and improves for all five families. Its exact two-sided
sign-flip p-value is also 0.0625. This conditional gain concerns
classification after acceptance; it does not establish that the score
separates unknown families.

\subsection{Validation Thresholds Accept Nearly All Unknowns}

At the known-validation threshold, residual community rejects 4.48\% of
held-out-family samples, compared with 3.92\% for generic topology. The
paired rejection increase is 0.56 percentage points
[0.01, 1.09] and is positive for four families, but the absolute operating
point remains poor: 95.52\% of unknown samples are accepted. Known acceptance
is 94.93\%, close to the intended 95\%, so the failure is not a missed
known-acceptance target. Instead, the known- and unknown-score distributions
overlap at the validation-selected boundary. The misses are not concentrated
near that boundary: accepted unknowns lie 1.093 score units below the
residual threshold on average versus 0.839 for generic topology, and the
mean cell-wise maximum margin is 1.608 versus 1.270. Residual community moves
only a small subset across the boundary while leaving most misses deep inside
the acceptance region.

The shared classifier is also weak: MSP-based known macro F1 is 0.2479, and
the residual-community prototype obtains 0.2143. This limits operational
interpretation because OSCR requires both correct known classification and
unknown rejection. It does not explain away the controlled comparison,
however: all methods use the same encoder, and classifier uncertainty still
achieves higher AUROC, lower FPR@95TPR, and higher OSCR.

\subsection{Decision Transitions Localize the Difference}

The decision-transition audit in Table~\ref{tab:transitions} pools 26,478
unknown-sample predictions from the 15 family--seed cells. The counts
represent model decisions rather than independent graph draws because each
test graph is evaluated by three seed-specific models. We therefore use this
audit descriptively; family-level sign patterns and exact tests bound the
small-sample interpretation.

\begin{table}[t]
\caption{Unknown-sample decision transitions at the known-validation
threshold. The two structural scores agree on 95.86\% of pooled
family--seed predictions, mostly by accepting the unknown sample.}
\label{tab:transitions}
\centering
\small
\setlength{\tabcolsep}{5pt}
\begin{tabular}{lrr}
\toprule
Decision pair & Count & Fraction \\
\midrule
Both accept & 24,808 & 93.69\% \\
Both reject & 573 & 2.16\% \\
Residual rejects only & 613 & 2.32\% \\
Generic rejects only & 484 & 1.83\% \\
\bottomrule
\end{tabular}
\end{table}

Residual-only rejections exceed generic-only rejections by 129 predictions,
or 0.49 percentage points in this sample-weighted pool. The direction is not
uniform across families. Residual-only rejection is more frequent in four
families, while WannaCry\_Worm has 86 residual-only and 115 generic-only
rejections. Winwebsec contributes 290 of the 613 residual-only cases. The
small aggregate increase thus comes from limited, family-concentrated
movements across the threshold rather than broad separation of unknown
samples from known support.

\begin{table}[t]
\caption{Known-sample transitions for predictions that are both accepted
and assigned to the correct family. Counts pool 41,133 family--seed
predictions. Residual community improves macro F1 without increasing the
total number of accepted-and-correct decisions.}
\label{tab:known-transitions}
\centering
\small
\setlength{\tabcolsep}{4pt}
\begin{tabular}{lrr}
\toprule
Decision pair & Count & Fraction \\
\midrule
Both correct & 9,488 & 23.07\% \\
Neither correct & 27,366 & 66.53\% \\
Residual correct only & 1,925 & 4.68\% \\
Generic correct only & 2,354 & 5.72\% \\
\bottomrule
\end{tabular}
\end{table}

Known-sample transitions provide a complementary view of the accepted-known
macro F1 result. The two scores agree on whether a sample is accepted and
correct in 89.60\% of pooled decisions. Generic-only correct decisions exceed
residual-only decisions by 429, or 1.04 percentage points. This does not
conflict with the positive macro F1 effect: macro F1 gives equal weight to
class-specific precision and recall, whereas the transition total weights
every sample equally. The residual block therefore redistributes known-class
errors without increasing the total number of accepted-and-correct
predictions.

The redistribution is also family-dependent. Residual-only correct decisions
exceed generic-only decisions for winwebsec (406 versus 320), but the generic
control has more in BazaLoader (509 versus 361), Mirai Botnet (476 versus
425), WannaCry\_Worm (459 versus 369), and Zeus Trojan (590 versus 364).
Together, the known and unknown transition audits show that the feature block
changes a limited set of decisions and that those changes remain concentrated
by held-out family.

\subsection{Scale Dependence and Family Reversals}

The residual-community score remains associated with structural scale after
training-set residualization. On known samples, its mean Spearman
correlations with node count, directed-edge count, and community count are
$-0.405$, $-0.381$, and $-0.406$; on unknown samples they are $-0.410$,
$-0.391$, and $-0.402$. The generic-topology control has substantially
smaller absolute correlations, ranging from 0.021 to 0.060 on known samples
and 0.127 to 0.148 on unknown samples. Residualization removes a linear
training-set prediction, but the resulting prototype distance still tracks
scale under held-out-family shift.

Size-stratified FPR@95TPR sharpens this pattern. Residual community improves
over generic topology by 0.0397 in the smallest node-count quartile, then
worsens by 0.0616, 0.0215, and 0.0235 in the next three quartiles. These
observations do not identify graph size as the sole cause of acceptance, but
they show that the added coordinates are not scale-neutral in the tested
score. Together with the winwebsec--WannaCry reversal, they bound the result:
the representation has family-specific effects rather than a stable
unknown-family signature.

\section{Discussion}

\subsection{Classification and Rejection Use Different Geometry}

Nearest-prototype classification depends on which class centroid is closest,
whereas rejection depends on the absolute distance to that closest centroid.
An added block can therefore improve accepted-known assignments without
moving unknowns outside known support. Residual community improves
accepted-known macro F1 in every family, yet AUROC, AUPRC, and OSCR are not
stable across families and FPR@95TPR deteriorates in all five. Both unanimous
directions have exact two-sided p-value 0.0625, the finest resolution
available here. The isotropic Euclidean score has no objective that reserves
an unknown margin; its F1 gain is evidence about conditional classification,
not open-set separation.

\subsection{Two Operating Points Reveal Score Overlap}

FPR@95TPR is a diagnostic operating point selected with test labels: the
residual-community score must falsely reject 91.60\% of known samples to
recover 95\% of unknown samples. The deployable threshold is selected only
from known validation data. It retains 94.93\% known acceptance but rejects
only 4.48\% of unknown samples. The score therefore provides no useful
trade-off near either high unknown recall or high known acceptance.

AUROC averages ranking over every threshold and can improve when a subset of
unknown samples moves upward, even if the operational boundary remains
inside heavy known--unknown overlap. OSCR also requires correct known
classification and degrades on average~\cite{dhamija2018reducing}. The
winwebsec--WannaCry reversal further shows why a pooled mean is insufficient:
a deployment encounters one unseen family, while the five-family analysis
remains descriptive.

\subsection{Residualization Does Not Imply Scale Invariance}

The ridge residual removes the linear component of standardized community
features predicted by generic topology in known-family training data. It
does not remove nonlinear dependence or held-out-family shift, and the
minimum-distance operation can restore scale association. This matters for
Louvain summaries because modularity resolution depends on network size and
interconnection~\cite{fortunato2007resolution}. Correlations near $-0.4$ and
deterioration in the three larger node-count quartiles indicate remaining
scale sensitivity, not a causal size mechanism. The dimension-matched
generic control therefore remains essential.

\subsection{Designing the Next Test}

A stronger test should separate representation quality, calibration, and
incremental structural information. It should first establish a stronger
known-family representation, using rejection-aware metric learning or
contrastive pretraining~\cite{wu2024contrastive}. A nested protocol can use
development-only unknown families for calibration while preserving distinct
final families. Community effects should also be tested in node-count and
density-matched strata, with family-level endpoints retained.

\section{Limitations}

The study covers structural FCGs, five held-out families, and one GIN
encoder; richer program semantics or other representations may yield
different score geometry. Five family units limit exact two-sided inference
to p-values no smaller than 0.0625. Cleaning ambiguous and duplicate graphs
changes the source distribution, two cells reach the 80-epoch cap, and
score--size correlations are diagnostic rather than causal.

\section{Conclusion}

We tested whether residual Louvain-community summaries add open-set
information beyond a GIN embedding and dimension-matched generic topology in
FCG-MFD. They do not provide stable held-out-family rejection in this tested
setting. AUROC
and AUPRC effects remain uncertain across held-out families, FPR@95TPR
worsens for all five family means, OSCR does not improve, and the
validation-fitted threshold accepts 95.52\% of unknown samples. The
consistent accepted-known F1 gain shows that the added coordinates can
refine classification among accepted examples without creating an unknown
margin. The score also remains associated with structural scale, and its
effect reverses between winwebsec and WannaCry\_Worm. In this GIN/FCG-MFD
study, community-enriched prototypes change known-class geometry without
yielding stable open-set separation. Graph OSR evaluations should pair
structural features with dimension-matched topology controls, operational
thresholds, and held-out-family analysis.

\bibliographystyle{IEEEtran}
\bibliography{references}

\end{document}